\documentstyle[epsfig,eqsecnum,aps]{revtex}
\begin{document}

\title{Spin-isospin excitations and $\beta ^+/EC$ half-lives of medium-mass
deformed nuclei}
\author{P. Sarriguren, E. Moya de Guerra, A. Escuderos}
\address{
Instituto de Estructura de la Materia,
Consejo Superior de Investigaciones Cient\'{\i }ficas,
Serrano 123, E-28006 Madrid, Spain }
\maketitle
\begin{abstract}
A selfconsistent approach based on a deformed HF+BCS+QRPA method
with density-dependent Skyrme forces is used to describe $\beta^+$-decay
properties in even-even deformed proton rich nuclei. Residual spin-isospin
forces are included in the particle-hole and particle-particle channels.
The quasiparticle basis contains neutron-neutron and proton-proton pairing 
correlations in the BCS approach, while neutron-proton pairing interaction
is treated as a residual force in QRPA.
We discuss the sensitivity of Gamow-Teller strength distributions and 
$\beta^+/EC$ half-lives to deformation, pairing and the strength of the
particle-particle interaction. The dependence on deformation is also compared
to that of spin $M1$ strength distributions.
\end{abstract}

\section{Introduction}

Exploring the properties of nuclei with unusual proton to neutron ratios is 
nowadays one of the major tasks and most active topics in nuclear structure
physics, both theoretical and experimentally \cite{review}. In this context, 
the beta-decay
properties of these exotic objects are of extreme importance in two different 
aspects. First, most nuclei of astrophysical interest \cite{astro} are those 
far from 
stability and their beta-decay rates have to be estimated theoretically.
Second, beta-decay properties and nuclear structure are intimately related.
It is clear that a precise and  reliable description of the ground state of the
parent nucleus and of the states populated in the daughter nucleus is
necessary to obtain a good description of the $\beta -$strength
distribution, and vice versa, failures to describe such distributions would
indicate that an improvement of the theoretical formalism is needed.

Microscopic models to describe the $\beta $-strength based on spherical
single-particle wave functions and energies with pairing and residual 
interactions treated in Random Phase Approximation (RPA) or Quasiparticle
Random Phase Approximation (QRPA) were first studied in Ref. \cite{sph}. Extensions
of these models to deal with deformed nuclei were done in Ref. \cite{kru}, where
a Nilsson potential was used to generate single-particle orbitals. Subsequent
extensions including Woods-Saxon type potentials \cite{moll}, residual 
interactions in the particle-particle channel \cite{hir}, consistent Hartree-Fock
(HF) mean field with residual interactions treated in Tamm Dancoff approximation
\cite{frisk}, selfconsistent approaches in spherical neutron-rich nuclei
\cite{doba} and based on an energy-density functional \cite {borzov}, can be 
also found in the literature.

In a previous work \cite{sarr1,sarr2} we studied ground state and 
$\beta $-decay properties of exotic nuclei on the basis of a deformed 
selfconsistent HF+BCS+QRPA calculation with density dependent effective 
interactions of Skyrme type.
This is a well founded approach that has been very successful in the 
description of spherical and deformed nuclei within the valley of 
stability \cite{flocard}.
 In this method once the parameters of the effective Skyrme interaction are 
determined, basically  by fits
to global properties in spherical nuclei over the nuclear chart, and the gap
parameters of the pairing force are specified, there are no free parameters
left. Both the residual interaction and the mean field are consistently obtained
from the same 2-body force. 
This is therefore a reliable method, suitable for extrapolations into the 
unstable regions approaching the drip lines. It is worth investigating
whether these powerful tools designed to account for the properties of
stable nuclei are still valid when approaching the drip lines.

The residual interaction  introduced in Refs. \cite{sarr1,sarr2}, consistent 
with the mean field, acts in the particle-hole channel ($ph$). However, 
it has been pointed out (see for instance Ref. \cite{kpp} and refs. therein) 
that for  a complete description of the $\beta ^{+}$ and $\beta \beta $ strengths, 
the inclusion of the particle-particle ($pp$) residual interaction is required.
We gave in \cite{sarr2} an example of the sensitivity of the results to the
inclusion of such a force but  a complete calculation within our formalism 
including this force is still missing. 
One of the purposes of this work is to extend our previous calculations by 
including a residual separable force in the $pp$ channel and to study
its effect in a systematic way to distinguish what is general and what is 
particular in the behavior of the nuclear response. To this end we include
in the QRPA calculation a 
separable neutron-proton pairing force as part of the residual interaction.

Following the same criteria as in our previous work \cite{sarr2}, we apply this
formalism to the study of a series of proton rich isotopes approaching
$N=Z$ in the mass region $A=70$. 
There are several reasons why this is a region of special interest to study 
$\beta $-decay. One is that the $Q$-value of the decay ($Q_{EC}$) is quite 
large in proton rich nuclei \cite{hamasag}. This means that a large fraction 
of the 
Gamow-Teller (GT) strength is contained within the $Q_{EC}$ window. Therefore, 
one can investigate directly the structure of the GT strength distribution by 
$\beta $-decay measurements without dealing with other indirect methods to 
extract the GT strength such as $(p,n)$ or $(n,p)$  charge exchange reactions.
Another aspect to stress is that the mass region $A\simeq 70$ is  at the border 
or beyond the scope of the full shell model calculations.
Predictions for the strength distributions, half-lives, and other decay 
properties obtained from the present formalism are of special 
relevance in this mass region since they may be at the moment the most 
reliable calculations. The present results can then be used as a guidance to 
further experimental 
searches. It is also worth mentioning that this mass region is known to present
particular structure effects, such as shape coexistence, which are the result
of a delicate balance of competing nuclear shapes \cite{frisk,petro}. 
It is then important to explore whether the dependence of the 
beta-decay properties on the deformation of the parent nucleus could be used
as an additional piece of information to elucidate the shape of the nucleus.
As a last point, we mention that, by  approaching systematically the $N=Z$ 
isotopes in various isotopic 
chains, we are placed in optimum conditions to observe whether agreement 
with experiment breaks down at some point as we approach $N=Z$. These isotopic
chains are a laboratory to test the validity of our formalism and to look for
failures related to the appearance of new physics such as neutron-proton 
pairing correlations not taken into account explicitly in the present mean field 
calculations, that may be important in $N=Z$ nuclei.
In  this work we extend the study of Refs. \cite{sarr1,sarr2} by
including a residual neutron-proton pairing interaction in the $J^\pi=1^+$ channel,
by studying the influence of usual BCS pairing correlations in the $N=Z$ isotopes,
and by discussing the similarities between the GT strength distributions and the
spin $M1$ strength distributions, which are the $\Delta T_z=0$ isospin counterparts
of the $\Delta T_z=\pm1$ GT transitions. 

In Section 2, we present a brief summary containing the basic points in our
theoretical description. Section 3 contains the results obtained for the GT 
strength distributions with a discussion of the dependence on the residual 
interaction, pairing correlations, and deformation. We also compare our results 
to the experimental available information. Spin $M1$ strength distributions
are studied in Section 4 discussing their analogies with the GT
strength distributions. The conclusions are given in Section V.

\section{Theoretical Approach}

In this Section we summarize briefly the theory involved in the microscopic
calculations presented in the next Sections. More details can be found in
Refs. \cite{sarr1,sarr2}. Our method consists in a selfconsistent
formalism based on a deformed Hartree-Fock  mean field obtained with a
Skyrme interaction including pairing correlations in the BCS approximation.
We consider in this paper the force SG2 \cite{giai} of Van Giai and Sagawa,
that has been successfully tested against spin and isospin excitations in 
spherical \cite{giai} and deformed nuclei \cite{sarr3}.
Comparison to calculations obtained with other Skyrme forces have been made
in Refs. \cite{sarr1,sarr2}, showing that the results do not differ in a 
significant way.
The single particle energies, wave functions, and occupations probabilities 
are generated from this mean field. 

For the solution of the HF equations we follow the McMaster procedure that
is based in the formalism developed in Ref.\cite{vautherin} as described in
Ref.\cite{vallieres}. Time reversal and axial symmetry are assumed.
The single-particle wave functions are expanded in
terms of the eigenstates of an axially symmetric harmonic oscillator in
cylindrical coordinates. We use eleven major shells. The method also
includes pairing between like nucleons in the BCS approximation with fixed
gap parameters for protons $\Delta _{\pi},$ and neutrons $\Delta _{\nu}$, which
are determined phenomenologically from the odd-even mass differences through
a symmetric five term formula involving the experimental binding energies 
\cite{audi}. The values used in this work are the same as those given in Ref.
\cite{sarr2}.

In a previous work \cite{sarr2} we analyzed the energy surfaces as a function 
of deformation for all the isotopes under study here. For that purpose, we 
performed constrained
HF calculations with a quadrupole constraint \cite{constraint} and we 
minimized the HF energy under the constraint of keeping fixed the nuclear 
deformation. Calculations in this paper are performed for the equilibrium 
shapes of each nucleus, that is, for the solutions, in general deformed, for 
which we obtained minima in the energy surfaces. Most of these nuclei present
oblate and prolate equilibrium shapes \cite{sarr2} that are very close in energy.

We add to the mean field a spin-isospin residual
interaction, which is expected to be the most important residual interaction
to describe GT transitions. This interaction contains two parts.
The particle-hole part is responsible 
for the position and structure of the GT resonance \cite{hir,sarr2} and
is derived selfconsistently from the same energy density functional (and Skyrme 
interaction) as the HF equation, in terms of the second derivatives of
the energy density functional with respect to the one-body densities 
\cite{bertsch}. The residual interaction is finally written in a separable
form by averaging the Landau-Migdal resulting force over the nuclear volume,

\begin{equation}
V^{ph}_{GT} = 2\chi ^{ph}_{GT} \sum_{K=0,\pm 1} (-1)^K \beta ^+_K 
\beta ^-_{-K} \, ,
\label{vph}
\end{equation}
where 
\begin{equation}
\beta ^+_K = \sum_{\pi\nu } \left\langle \nu | \sigma _K |\pi \right\rangle 
a^+_\nu a_\pi \, .
\end{equation}
The coupling strength is given by \cite{sarr1,sarr2}

\begin{equation}
\chi ^{ph}_{GT}=-\frac{3}{8\pi R^{3}} \left\{ 
t_{0}+\frac{1}{2}k_{F}^{2}\left( t_{1}-t_{2}\right) +\frac{1}{6}t_{3}
\rho ^{\alpha }\right\} \label{chiph}
\end{equation}
and it is completely determined by the Skyrme parameters 
$t_0,t_1,t_2,t_3,\alpha $, the nuclear radius $R$, and the Fermi 
momentum $k_F$, obtained with the same Skyrme force for the given nucleus.

The particle-particle part is a neutron-proton pairing force in the $J^\pi=1^+$
coupling channel.
We introduce this interaction in the usual way \cite{hir,kpp,muto}, that is, 
in terms of a 
separable force with a coupling constant $\kappa ^{pp}_{GT}$, which is fitted 
to the phenomenology,

\begin{equation}
V^{pp}_{GT} = -2\kappa ^{pp}_{GT} \sum_K (-1)^K P ^+_K P_{-K} \, ,
\label{vpp}
\end{equation}
where
\begin{equation}
P ^+_K = \sum_{\pi\nu} \left\langle \pi \left| \left( \sigma_K\right)^+ \right|\nu 
\right\rangle  a^+_\nu a^+_{\bar{\pi}} \, .
\end{equation}

The two forces $ph$ and $pp$ in Eqs. (\ref{vph}) and (\ref{vpp}) 
are defined with a positive and a negative sign, respectively, according to their 
repulsive and attractive character, so 
that the coupling strengths $\chi$ and $\kappa $ take positive values.

The peak of the
GT resonance is almost insensitive to the $pp$ force and $\kappa ^{pp}_{GT}$ 
is usually adjusted to reproduce the half-lives \cite{hir}.
However, one should be careful with the choice of this coupling constant.
Since the $pp$ force is introduced independently of the mean field, 
if $\kappa ^{pp}_{GT}$ is strong enough it may happen that the QRPA
collapses, because the condition that the ground state be stable against the
corresponding mode is not fulfilled. This happens because the $pp$ force, being 
an attractive force, makes the 
GT strength to be pushed down to lower energies with increasing values of 
$\kappa ^{pp}_{GT}$. 

A careful search of the optimal strength can certainly be done for each 
particular case, but this is not our purpose here. Instead, we have chosen
the same coupling constant for all nuclei considered here. This value has been
obtained under the requirements of improving in general the agreement with
the experimental half-lives while being still far from the values leading
to the collapse. This will be discussed in more detail in the next Section.

The proton-neutron  quasiparticle random phase approximation (QRPA) 
phonon operator for GT excitations in
even-even nuclei is written as

\begin{equation}
\Gamma _{\omega _{K}}^{+}=\sum_{\pi\nu}\left[ X_{\pi\nu}^{\omega _{K}}\alpha
_{\nu}^{+}\alpha _{\bar{\pi}}^{+}+Y_{\pi\nu}^{\omega _{K}}\alpha _{\bar{\nu}}\alpha
_{\pi}\right]\, ,  \label{phon}
\end{equation}
where $\alpha ^{+}\left( \alpha \right) $ are quasiparticle creation
(annihilation) operators, $\omega _{K}$ are the excitation energies, and 
$X_{\pi\nu}^{\omega _{K}},Y_{\pi\nu}^{\omega _{K}}$ the forward and backward
amplitudes, respectively. 
The equations of motion are solved in the proton-neutron QRPA.

From the QRPA equations the forward and backward amplitudes are obtained as 
\cite{muto}
\begin{equation}
X_{\pi\nu}^{\omega _{K}}=\frac{1}{\omega _{K}-\varepsilon_{\pi\nu}}
\left[ 2\chi ^{ph}_{GT} \left( a_{\pi\nu}M_{-}^{\omega _{K}}+b_{\pi\nu}
M_{+}^{\omega _{K}}\right)
-2\kappa ^{pp}_{GT} \left( c_{\pi\nu}M_{--}^{\omega _{K}}+d_{\pi\nu}
M_{++}^{\omega _{K}}\right) \right] \, ,
  \label{xwa}
\end{equation}

\begin{equation}
Y_{\pi\nu}^{\omega _{K}}=\frac{1}{\omega _{K}+\varepsilon_{\pi\nu}}
\left[ 2\chi ^{ph}_{GT} \left( a_{\pi\nu}M_{+}^{\omega _{K}}+b_{\pi\nu}
M_{-}^{\omega _{K}}\right)
+2\kappa ^{pp}_{GT} \left( c_{\pi\nu}M_{++}^{\omega _{K}}+d_{\pi\nu}
M_{--}^{\omega _{K}}\right) \right] \, ,
\label{ywa}
\end{equation}
with $\varepsilon_{\pi\nu}=E_{\nu}+E_{\pi}$ the two-quasiparticle excitation
energy in terms of the quasiparticle energies $E_{i}$. $M^{\omega
_{K}}$ are given by

\begin{equation}
M_{-}^{\omega _{K}}=\sum_{\pi\nu}\left( a_{\pi\nu}X_{\pi\nu}^{\omega _{K}}-
b_{\pi\nu}Y_{\pi\nu}^{\omega _{K}}\right) \, ,
\label{mplus}
\end{equation}

\begin{equation}
M_{+}^{\omega _{K}}=\sum_{\pi\nu}\left( b_{\pi\nu}X_{\pi\nu}^{\omega _{K}}-
a_{\pi\nu}Y_{\pi\nu}^{\omega _{K}}\right) \, ,
\label{mminus}
\end{equation}

\begin{equation}
M_{--}^{\omega _{K}}=\sum_{\pi\nu}\left( c_{\pi\nu}X_{\pi\nu}^{\omega _{K}}+
d_{\pi\nu}Y_{\pi\nu}^{\omega _{K}}\right) \, ,
\label{mmminus}
\end{equation}

\begin{equation}
M_{++}^{\omega _{K}}=\sum_{\pi\nu}\left( d_{\pi\nu}X_{\pi\nu}^{\omega _{K}}+
c_{\pi\nu}Y_{\pi\nu}^{\omega _{K}}\right) \, ,
\label{mmplus}
\end{equation}
with
\begin{equation} 
a_{\pi\nu}=v_{\nu}u_{\pi}\Sigma _{K}^{\nu\pi};\ \ \ 
b_{\pi\nu}=u_{\nu}v_{\pi}\Sigma _{K}^{\nu\pi };\ \ \
c_{\pi\nu}=u_{\nu}u_{\pi}\Sigma _{K}^{\nu\pi};\ \ \ 
d_{\pi\nu}=v_{\nu}v_{\pi}\Sigma _{K}^{\nu\pi}\, , 
\end{equation}
where $v'$s are occupation amplitudes ($u^2=1-v^2$) and $\Sigma _{K}^{\nu\pi}$ 
spin matrix elements connecting neutron and proton states with spin operators
\begin{equation}
\Sigma _{K}^{\nu\pi}=\left\langle \nu\left| \sigma _{K}\right| \pi\right\rangle 
\, .
\label{sigma}
\end{equation}

The solution of the QRPA equations can be found solving first a dispersion
relation, which is now of fourth order in the excitation energies $\omega$. 
Then, for each value of the energy the amplitudes $M$ are determined
by using the normalization condition of the phonon amplitudes

\begin{equation}
\sum_{\pi\nu} \left[ \left( X_{\pi\nu}^{\omega _K}\right) ^2- 
\left( Y_{\pi\nu}^{\omega _K}\right) ^2 \right] =1 \, .
\end{equation}
The technical procedure to solve these QRPA equations is well described in Ref.
\cite{muto}.

In the intrinsic frame the GT transition amplitudes connecting the
QRPA ground state 
$\left| 0\right\rangle \ \ \left( \Gamma _{\omega _{K}} \left| 0\right\rangle 
=0 \right)$ to one phonon states $\left| \omega _K \right\rangle \ \ \left( 
\Gamma ^+ _{\omega _{K}} \left| 0\right\rangle = \left|
\omega _K \right\rangle \right)$, are given by
\begin{equation}
\left\langle \omega _K | \sigma _K t^{\pm} | 0 \right\rangle = \mp 
M^{\omega _K}_\pm \, .
\end{equation}

The Ikeda sum rule is always fulfilled in our calculations

\begin{equation}
\sum_\omega \left[ \left( M^{\omega }_{-}\right) ^2-
\left( M^{\omega }_{+}\right) ^2 \right] = 3(N-Z) \, .
\end{equation}

\section{Decay properties}

The bulk properties of the isotope chains considered here were already 
studied in Ref. \cite{sarr2}. Binding energies, charge radii, quadrupole
moments, and moments of inertia were discussed and compared successfully
to available 
experimental data. Here we will concentrate on the decay properties.

In this Section we discuss the results obtained for the GT strength
distributions, half-lives, and summed strengths in the proton rich isotopes 
belonging to the $A\simeq 70$ mass region. The results correspond to QRPA
calculations with the Skyrme force SG2 and they have been performed for the nuclear
shapes that minimize the HF energy. 

Figures showing the GT strength distributions are plotted versus
the excitation energy of daughter nucleus.
The distributions of the GT strength have been folded with $\Gamma =1$
MeV width Gaussians to facilitate the comparison among the various calculations,
so that the original discrete spectrum is transformed into a continuous profile.
These distributions are given in units of $g_A^2/4\pi$ and one should keep in
mind that a quenching of the $g_A$ factor, typically 
$g_{A,eff}=(0.7-0.8)\  g_{A,free}$
is expected on the basis of the observed quenching in charge exchange reactions 
and spin $M1$ transitions in stable nuclei, where $g_{s,eff}$ is also known
to be approximately $0.7\ g_{s,free}$. Therefore, a reduction factor of about two
is expected in these strength distributions in order to compare with experiment.
This factor is of course taken into account when comparing to experimental
half-lives.

\subsection{The role of the residual interactions}

Figure 1 illustrates the effect of the residual interactions on the uncorrelated
2-quasiparticle calculation.
The calculations are done for the oblate and prolate shapes of $^{74}$Kr.
The coupling strength of the $ph$ residual interaction $\chi ^{ph}_{GT}$ is 
obtained
from Eq. (\ref{chiph}), and its value for $A=74$ and Skyrme force SG2 is 
$\chi ^{ph}_{GT}=0.37$ MeV. The coupling strength of the $pp$ residual interaction
is varied from  $\kappa ^{pp}_{GT}= 0$ to  $\kappa ^{pp}_{GT}= 0.07$ MeV.

If we compare first the calculation with only the $ph$ residual interaction 
(dotted line) to the uncorrelated one (thin solid line), we can see that the
repulsive $ph$ force introduces two types of effects. One is a shift of the 
GT strength to higher excitation energies with the corresponding displacement 
of the position of the GT resonance. The other is a reduction of the total GT 
strength. Obviously these effects are more pronounced as we increase the
value of the coupling strength $\chi ^{ph}_{GT}$. 
If we now introduce a $pp$ residual interaction (dashed and solid lines), 
we can see
that its effect, being an attractive force, is to shift the strength to
lower excitation energies, reducing the total GT strength as well. The shift
and reduction of strength is more pronounced at high excitation energies
and the position of the GT resonance is hardly affected by this interaction.
The GT strength is also pushed a bit by the $pp$ interaction to lower energies 
in the low energy excitation region. This effect, although small, is of great 
relevance in 
the calculation of the $\beta ^+/EC$ half-lives, which are only sensitive to 
the distribution of the strength contained in the energy region below the 
$Q_{EC}$ window. By comparing the curves obtained from  $\kappa ^{pp}_{GT}= 0.05$
MeV (dashed line) and  $\kappa ^{pp}_{GT}= 0.07$ MeV (solid line), we can also 
see how these effects are more pronounced as we increase the value of the 
coupling strength.

This is the reason why the usual procedure to fit these two coupling strengths
$\chi ^{ph}_{GT}$ and $\kappa ^{pp}_{GT}$ is as follows: First one chooses 
$\chi ^{ph}_{GT}$ 
to reproduce the position of the GT resonance, usually determined from charge 
exchange reactions $(p,n)$ and $(n,p)$, and then one chooses $\kappa ^{pp}_{GT}$ to
reproduce the half-life.
In our work, since the value of $\chi ^{ph}_{GT}$ is determined selfconsistently 
and the known GT resonances are reasonably well described \cite{sarr2}, we do 
not carry 
out this case by case fitting procedure. However, to determine the value of 
$\kappa ^{pp}_{GT}$, 
we calculate first GT strength distributions and
half-lives and compare the latter with experiment to extract a value that
producing a reasonable agreement is still within the range of values 
compatible with the correct treatment of the QRPA.

In Figure 2 we can see the result of the calculation of the half-lives
in $^{76}$Sr as a function of the coupling strength  $\kappa ^{pp}_{GT}$. The 
half-lives are obtained from the familiar expression,

\begin{equation}
T_{1/2}^{-1}=\frac{A ^{2}}{D}\sum_{\omega }f\left( Z,\omega \right)
\left| \left\langle 1_{\omega }^{+}\left\| \beta ^{+}\right\|
0^{+}\right\rangle \right| ^{2} \, , \label{t12}
\end{equation}
where  $f\left( Z,\omega \right) $ are the Fermi integrals. 
Note that in our previous work \cite{sarr2}, we used for these integrals the
values tabulated in \cite{gove}. In the present calculations we compute the
Fermi integrals numerically for each $Z,\omega$ values. By this procedure we
get more accurate results.

We use $D=6200$ s and include standard effective factors

\begin{equation}
A^{2}=\left[ \left( g_{A}/g_{V}\right) _{eff}\right] ^{2}=\left[
0.77\left( g_{A}/g_{V}\right) _{free}\right] ^{2}=0.90 \, . \label{quen}
\end{equation}
 
The half-life decreases with increasing values of  $\kappa ^{pp}_{GT}$. 
This is clear
because as  $\kappa ^{pp}_{GT}$ increases the strength becomes more concentrated at
low excitation energies below $Q_{EC}$ and therefore the half-lives are smaller.
This is true up to values around  $\kappa ^{pp}_{GT} = 0.1$ MeV, where the collapse
of the QRPA takes place.
In the case of  $^{76}$Sr, we get an optimum value of  $\kappa ^{pp}_{GT}=0.03$ MeV
and  $\kappa ^{pp}_{GT} =0.07$ MeV for the oblate and prolate shapes, respectively.
This value will of course depend, among other factors, on the nucleus, shape, and 
Skyrme interaction and a case by case fitting procedure could be carried out.
Nevertheless, we have done calculations for other cases and found that, in
general, values around  $\kappa ^{pp}_{GT}=0.07$ MeV improve the agreement 
with experiment
in most cases and since this is a valid value far from collapse, we have chosen
$\kappa ^{pp}_{GT}=0.07$ MeV as the value of the coupling constant of the 
$pp$ residual 
interaction.

In Ref. \cite{hir} Homma and collaborators studied $\beta -$decay properties 
using
Nilsson+BCS+QRPA with $ph$ and $pp$ separable residual interactions. They
considered a wide range of nuclei to extract phenomenologically the coupling
strength of such residual interactions by fits to experimental half-lives.
They obtained the following dependence on the mass number: 
$\chi ^{ph}_{GT}=5.2/A^{0.7}$ MeV, $\kappa ^{pp}_{GT}=0.58/A^{0.7}$ MeV.
For $A=70$ this gives a $ph$ strength $\chi ^{ph}_{GT}=0.27$ MeV and a $pp$
strength $\kappa ^{pp}_{GT}=0.03$ MeV. The $ph$ strength is somewhat
smaller than the one obtained from Eq. (\ref{chiph}), derived selfconsistently
from our mean field. The $pp$ strength is also smaller than the value adopted
here. This is consistent with the trend observed in Fig. 1 of Homma {\it et al.}
\cite{hir} that shows that for decreasing values of  $\chi ^{ph}_{GT}$ one 
needs smaller values of  $\kappa ^{pp}_{GT}$.
On the other hand, a discussion of the adequacy of our $ph$ strength is also
demonstrated in Ref. \cite{sarr2}, where we compared our results on the 
position of the GT resonance with experimental data from $(p,n)$ and $(n,p)$
reactions in this mass region. The fact that our $ph$ and $pp$ strengths
are somewhat different from those in Ref. \cite{hir} is not surprising since
the mean fields are also different.

\subsection{The role of BCS correlations}

As already mentioned in the Introduction, our theoretical treatment does
not explicitly include neutron-proton  pairing correlations in the mean 
field. Thus, our quasiparticle basis only includes neutron-neutron and
proton-proton pairing correlations in the BCS approach.
In principle, one could extend the BCS treatment to include also neutron-proton
pairing correlations in the mean field. This may be important particularly
for N=Z nuclei.

In Ref. \cite{sarr2} we studied  bulk properties (binding energies, charge radii,
quadrupole moments, moments of inertia,...) of the nuclei considered here
and found that the agreement between theory and
experiment is as good for the $N=Z$ as for the $N=Z+2,Z+4,Z+6$ isotopes.
We concluded that the effect of neutron-proton pairing correlations 
on these bulk properties is roughly taken into account by the use of the
phenomenological gap parameters $\Delta _{\pi},\Delta _{\nu}.$ 

The main effect of taking into account neutron-proton pairing in the HF+BCS 
calculation would be to increase the diffuseness of the Fermi surface 
\cite{nppairing}. This diffuseness is proportional to
the gap parameters. It is therefore interesting to study the sensitivity of 
the GT strength to the gap parameters in the $N=Z$ nuclei $^{64}$Ge,  $^{68}$Se, 
$^{72}$Kr, and  $^{76}$Sr. To this end we compare in Fig. 3 the QRPA results 
obtained with the SG2 force in those nuclei for various values 
of the gap parameters differing by $\pm 0.5$ MeV from the values extracted
from the phenomenology. To make the discussion easier we did not include
the $pp$ residual interaction in those figures.

The main effect of the BCS correlations is to create new transitions
that were forbidden in the absence of such correlations. Since the occupation
probabilities are now different from 0 or 1, the already existing peaks at
$\Delta=0$ will decrease when $\Delta>0$, while new strength will appear at 
other energies and will increase with increasing gap 
parameters. This new strength is in general located at high excitation energy,
while the strength already present at $\Delta=0$ is mainly concentrated at 
lower energy.
As a consequence, the main effect of increasing the Fermi diffuseness is to 
smooth out the profile of the GT strength distribution, increasing the strength at
high energies and decreasing the strength at low energies.

It is clear from Fig. 3 that each particular case has its own peculiarities
but the general trend described above plays the major role.
It is also interesting to remark that in the high energy tail beyond the GT
resonance, the effects of increasing $\Delta$ and increasing  $\kappa ^{pp}_{GT}$
are opposite. As we mentioned before, increasing  $\kappa ^{pp}_{GT}$ tends to
deplete those tails.

\subsection{The role of deformation}

The role of deformation on the GT strength distributions can be summarized, 
in general, in two types of effects. First, deformation breaks down the 
degeneracy of the spherical shells and this implies that the GT strength
distributions corresponding to deformed nuclear shapes will be much more
fragmented than the corresponding spherical ones. Second, the energy levels
of deformed orbitals coming from different spherical shells cross each other
in a way that depends on the magnitude of the quadrupole deformation as well
as on the oblate or prolate character. This level crossing may lead in some
instances to similar profiles in the GT strength distributions of
the various coexisting nuclear shapes but in other cases it may lead to
sizeable differences between the GT strength distributions corresponding
to different shapes of the same parent nucleus. This fact can be exploited
to gain information on what is the nuclear shape of a nucleus by just looking
at the structure of its $\beta $-decay.

Fig. 4 summarizes the main results of this work. They correspond to the GT
strength distributions obtained from our HF+BCS+QRPA with SG2 in the isotope
chains of Ge, Se, Kr, and Sr. We can see in this figure 
the results obtained for the possible nuclear shapes of the isotopes
ranging from $N=Z$ to $N=Z+6$. As usual the strengths are in units of 
$g_A^2/4\pi$ and are plotted versus the excitation energy of the daughter
nucleus.

We can see that in general the $N=Z$ isotopes of each chain contain the
maximum strength as it corresponds to the most unstable nuclei. This strength
becomes smaller and smaller as we increase the number of neutrons approaching
the stable isotopes. The excitation energy of the GT resonance also
decreases with the number of neutrons.

If we look in Fig. 3 the case of Ge and Se isotopes, we can see that
there are no signatures of the actual character, oblate or prolate, of the 
deformation of the parent nucleus: Both equilibrium shapes
produce similar GT profiles independently of whether it is oblate or prolate. 
Then, it can be concluded that in these isotopes one cannot use GT distributions
to distinguish between the two coexistent shapes predicted by the theory. 
On the other hand, the figures corresponding to Kr and Sr isotopes,
show differences in the GT profiles of the various shapes for
each isotope that can be easily identified  even within the $Q_{EC}$ window.
A case by case analysis allows one to conclude that the most favorable candidates
to look for deformation effects based on $\beta $-decay measurements are
$^{74}$Kr and $^{76,78,80}$Sr.
In these isotopes the GT strengths within the $Q_{EC}$ window are different enough 
to distinguish between different equilibrium shapes. On the contrary, there are
cases like $^{72}$Kr where though the large $Q_{EC}$ value ($Q_{EC}=5$ MeV)
make it worth to investigate experimentally, the measured strength
may not be conclusive on the shape because 
oblate and prolate shapes produce similar profiles. Other isotopes
like $^{76,78}$Kr have small  $Q_{EC}$ values that will not allow a clear 
conclusion either, while
$^{74}$Kr seems to be the best Kr candidate. It has a $Q_{EC}$ window (3.1 MeV)
big enough to distinguish between a continuously increasing profile, as the prolate
shape predicts, or a completely developed bump structure peaked at around
1.5 MeV and vanishing at about the $Q_{EC}$ value, as the oblate shape predicts.
A similar situation happens in the case of  $^{76}$Sr ($Q_{EC}=6.1$ MeV). The
oblate shape produces a peak centered at around 1 MeV and a second one centered
at 6 MeV, while the prolate shape produces a GT distribution that increases 
almost continuously up to 6 MeV. The case of  $^{78}$Sr ($Q_{EC}=3.8$ MeV) is 
again a clear case where one can distinguish between the bump structure 
generated by the spherical shape or the continuously increasing pattern 
of the prolate case. The same is true for the   $^{80}$Sr isotope, although
in this case $Q_{EC}$ is not so big ($Q_{EC}=1.9$ MeV).

\subsection{Comparison to experiment}

In this subsection we compare to experiment our calculated $Q_{EC}$ and
total $\beta ^{+}/EC$ half-lives $T_{1/2}$, as well as a few cases where the 
GT strength distribution have been
partially measured by $\beta$-decay. We also discuss the GT strengths
contained within the $Q_{EC}$ windows.
The difference with respect to the values quoted in our previous work \cite{sarr2}
is that
now the $pp$ interaction is included. There is also a minor difference in the 
calculation of $Q_{EC}$,
\begin{equation}
Q_{EC}=m_{\pi}-m_{\nu}+m_{e}-\left( \lambda _{\nu}+E_{\nu}\right) _{\left(
N,Z\right) }+\left( \lambda _{\pi}-E_{\pi}\right) _{\left( N,Z-2\right) } \, .
\label{qec}
\end{equation}
The $Q_{EC}$ values in Tables 1-4 of Ref. \cite{sarr2} were calculated 
approximating the lowest quasiparticle
energies $E_\nu$ and $E_\pi$ by the gap parameters $\Delta_\nu$ and $\Delta_\pi$,
respectively, while $Q_{EC}$ values in Table 1 here are calculated using their
actual values
$E=\sqrt{\left( \epsilon -\lambda \right) ^{2}+\Delta ^{2}}$.

We can see in Table 1 these results obtained from  QRPA calculations with 
the Skyrme force SG2 and for the different shapes, oblate, prolate or spherical, 
where the minima occur for each isotope. We did not include the stable nuclei,
all having $Q_{EC}$ less than zero and infinite half-lives.
It should also be mentioned that the total GT strength below  $Q_{EC}$ has been
reduced by a quenching factor 0.6 to be consistent with the same quenching
used in the calculation of the half-lives in Eq. (\ref{t12}).

We get a very good agreement with the measured $Q_{EC}$ values in practically
all cases and the values obtained with the various shapes are quite similar.
The half-lives are also well reproduced in most cases. Only in the most 
stable isotopes, where the half-lives are very large, we find some noticeable
discrepancies but this is not so relevant because in these cases the 
$Q_{EC}$ values are very small and therefore, the half-lives are only sensitive
to a tiny region of the GT tail at low energies. 
In other cases the agreement is very reasonable.
The sums of the GT strength up to the  $Q_{EC}$ value do not differ much from
one shape to another although this does not mean that the structure of the
profiles are equivalent. As we have seen in the last subsection, there are 
cases,  $^{74}$Kr and $^{76,78,80}$Sr, whose profiles can be easily 
distinguished although the final summed strength is very similar.

Table 2 contains experimental information on GT summed strengths in the $N=Z$
nuclei  $^{72}$Kr \cite{exp72} and  $^{76}$Sr \cite{exp76} up to different 
values of the excitation energy, always below  $Q_{EC}$. They are compared
to our theoretical calculations for the two equilibrium shapes using the same
quenching factor as in Eq. (\ref{quen}).
From this comparison an oblate shape for  $^{72}$Kr and a prolate shape for
$^{76}$Sr seems to be favored, but this is not yet sufficient for a conclusive
answer. 

In any case it would be worth to extend the measurements in $^{76}$Sr
to higher excitation energies because it is a very good example where 
deformation effects are visible (see Fig. 4 for $^{76}$Sr). On the other hand, 
the extension of these measurements to the case of $^{72}$Kr may not be so 
relevant since one could not distinguish one shape from another (see Fig. 4
for $^{72}$Kr).

\section{Magnetic properties}

The spin $M1$ transitions are the $\Delta T_{z}=0$ isospin counterparts of
the $\Delta T_{z}=\pm 1$ GT transitions. The study carried out in 
\cite{sarr3} for several isotope chains showed that the $M1$ strength
depends on deformation, not only for orbital but also for spin excitations.
Thus, on general grounds, one may argue that the deformation dependence of
the GT strength discussed earlier should be comparable to that of the spin 
$M1$ strength considered in \cite{sarr3}.
This is expected to be particularly the case when $N=Z$.

In a similar way to Fig. 4 for the GT distributions, we show in Fig. 5
the profiles of the spin $M1$ strength distributions for the same isotope
chains. The results correspond to the selfconsistent
HF+BCS+QRPA calculations with the SG2 interaction. The details of these 
calculations are
described in \cite{sarr3} and closely follow those of the GT strength
except for the $\Delta T_{z}=0$ character of the $M1$ operator.

Similarly to the GT strength distributions in the Ge and Se isotope chains in 
Fig. 4, we can see that the $B_\sigma (M1)$ strength distributions for these
two isotopes in Fig. 5 have similar structures for the oblate and 
prolate shapes. They have a big resonance located at the same energy around 10
MeV and a small
bump at about 5 MeV. However, the strength contained in the prolate peak is
always larger than the oblate one, contrary to what happened with the GT
strengths that were comparable. The spherical shape in $^{74}$Se produces
much less strength than the deformed shapes.
We can also see in Fig. 5 for Kr and Sr isotopes  
that the profiles of the $M1$ strengths corresponding to the  prolate and 
oblate distributions can be clearly distinguished, similarly to the case of the
GT distributions in Fig. 4. The strength corresponding to the prolate 
shape is again the largest. Therefore, clear similarities between the GT and $M1$ 
strength distributions are observed, but some differences can be seen as well.
In particular, for spin $M1$ strength distributions the position and strength 
of the resonance
are  practically the same in all nuclei in a given isotope chain. 
This is different to what happened with the GT strength distributions,
where the  $\beta ^+$-GT strength decreases very fast with
increasing $N-Z$ due to Pauli blocking, while $M1$ strengths are not affected
by this.

\section{Conclusions}

We studied $\beta^+$-decay in various isotopic chains of medium-mass 
proton-rich nuclei within the framework of the selfconsistent deformed
HF+BCS+QRPA with Skyrme interactions. Our spin-isospin residual interaction
contains a particle-hole part, which is derived selfconsistently from the
Skyrme force, and a particle-particle part, which is a separable force 
representing a neutron-proton pairing force. Compared to previous calculations,
where the latter residual interaction was not considered, we obtain in general
better agreement with experiment for half-lives. It is worth mentioning that
for the N=Z isotopes this agreement is comparable to that achieved for the other 
isotopes with an excess of neutrons. This indicates that using the phenomenological
gap values for neutrons and protons, as well as the neutron-proton pairing
correlations as a residual force in QRPA, is sufficient to account at least
for this experimental information.

From our study of the dependence on the shape of the GT strength distributions
we conclude that there are some interesting cases that are worth to explore 
experimentally such as $^{74}$Kr and $^{76,78,80}$Sr. In these examples
the measured  $\beta^+$ strengths below $Q_{EC}$ could be used to identify 
their equilibrium shapes. In particular, we have observed in $^{76}$Sr various 
compatible indications pointing out towards the same conclusion: a strong
prolate component in the ground state that is suggested from the comparison to
experiment of half-lives and GT strength measured at low excitation energies.
Nevertheless, more experimental information is still needed to reach a 
conclusive answer.

We also studied the energy distributions of the spin $M1$ strength and the
similarities and differences with their GT counterparts. We conclude
that the main features of GT and spin $M1$ strength distributions are similar.
This suggests that we may learn about properties that are observable in highly
unstable nuclei (like $\beta $ decay) from properties that are observable in
stable nuclei (like $M1$'s), and vice versa.
Information on spin $M1$ excitations can be extracted from both leptonic
and hadronic probes. In stable nuclei, the combined analysis of electron,
photon, and proton scattering data has provided reliable information on
orbital and spin $M1$ strength distributions \cite{expm1}. In principle, 
the same type of experiments could be performed on the proton rich nuclei
considered here. Inelastic $(p,p')$ scattering experiments might be 
particularly suitable to explore the spin part of the $M1$ strength and
to test the predictions of the present paper.

\vskip 1cm

\begin{center}
{\Large \bf Acknowledgments} 
\end{center}
This work was supported by DGESIC (Spain) under contract number PB98-0676. 
One of us (A.E.) thanks Ministerio de Educaci\'{o}n y Cultura (Spain) for support.

\vfill\eject

\vfill \eject

\vfill\eject

\begin{figure}[t]
\epsfig{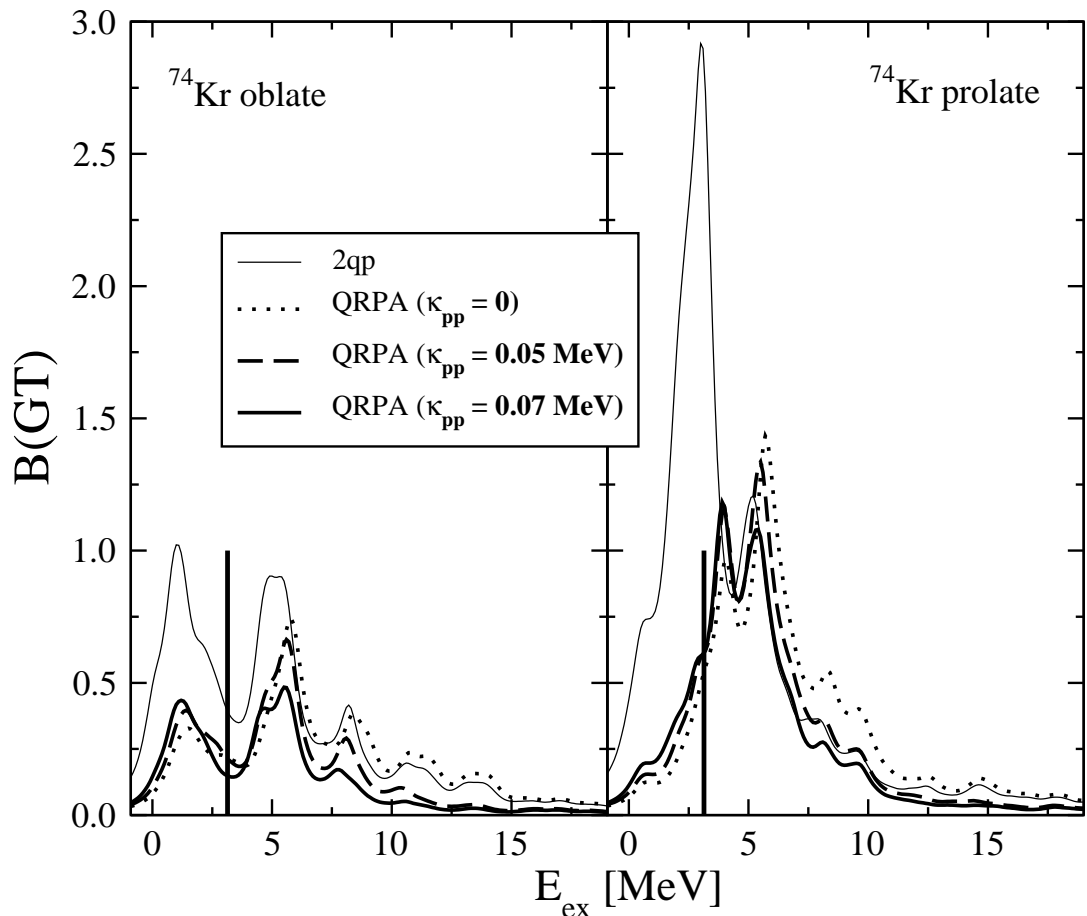}
\caption{GT strength distributions [$g_{A}^{2}/4\pi $] in $^{74}$Kr 
plotted versus the excitation energy of the daughter nucleus. Calculations are
done in QRPA with the force SG2 for various values of the coupling strength 
$\kappa ^{pp}_{GT}$ of the particle-particle force.
Vertical lines indicate experimental $Q_{EC}$ values
(see Table 1 for the theoretical $Q_{EC}$ values).}
\end{figure}

\vfill\eject

\begin{figure}[t]
\epsfig{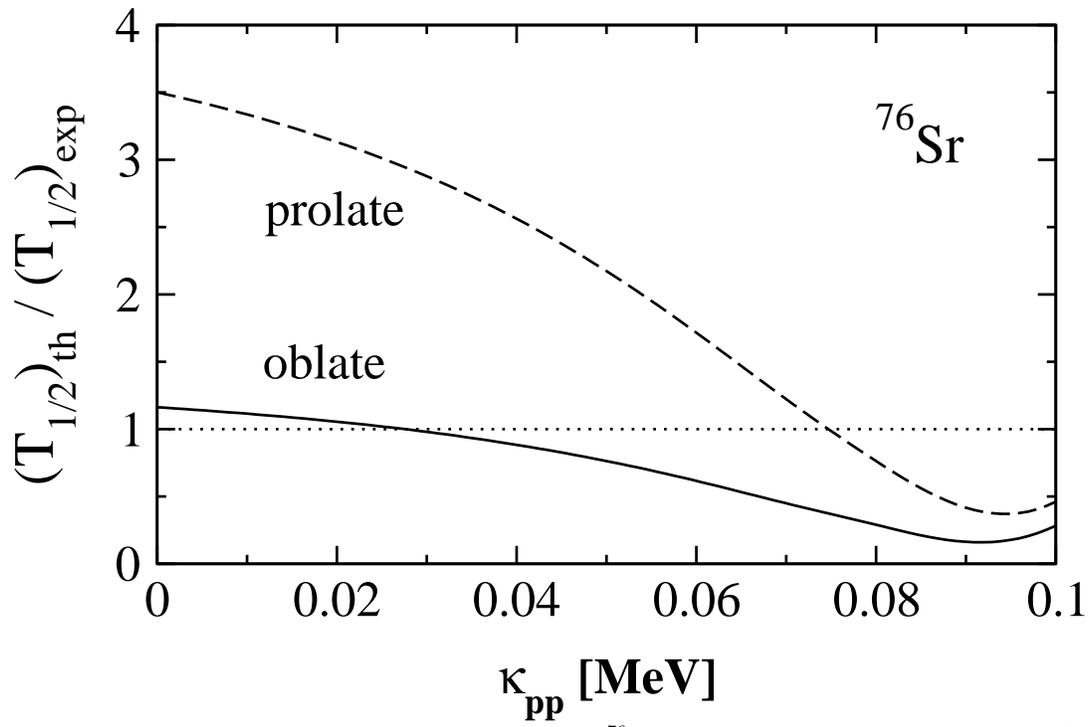}
\caption{Ratios of calculated to experimental half-lives in $^{76}$Sr 
as a function of the coupling strength $\kappa ^{pp}_{GT}$ of the 
particle-particle force.}
\end{figure}

\vfill\eject

\begin{figure}[t]
\epsfig{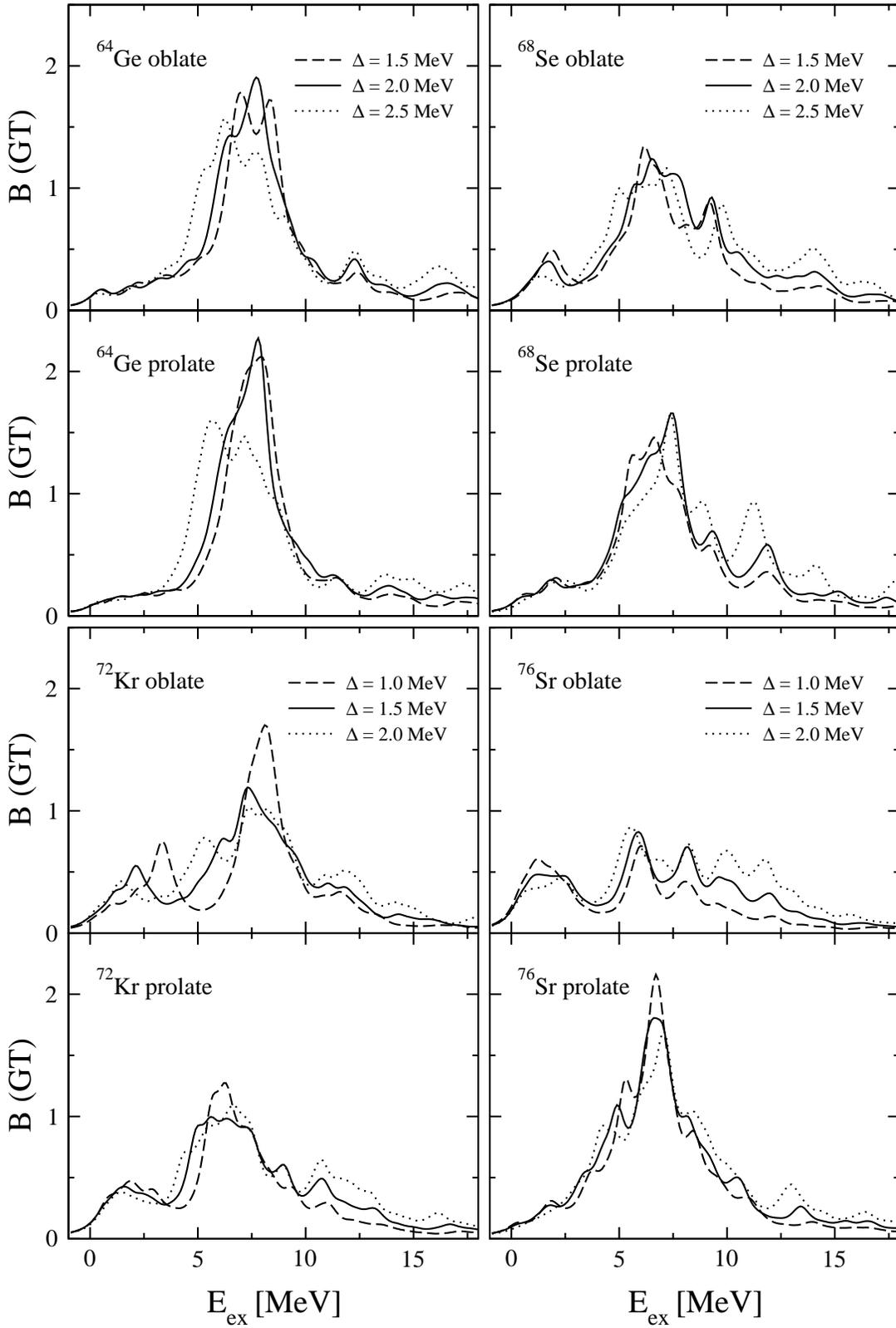}
\caption{Pairing effect in the QRPA Gamow-Teller strength distribution
in the $N=Z$ isotopes $^{64}$Ge, $^{68}$Se, $^{72}$Kr, and $^{76}$Sr.
Solid lines correspond to calculations with empirical pairing 
gaps, dotted (dashed) lines correspond to calculations with empirical gaps
plus (minus) 0.5 MeV.}
\end{figure}

\vfill\eject

\begin{figure}[t]
\epsfig{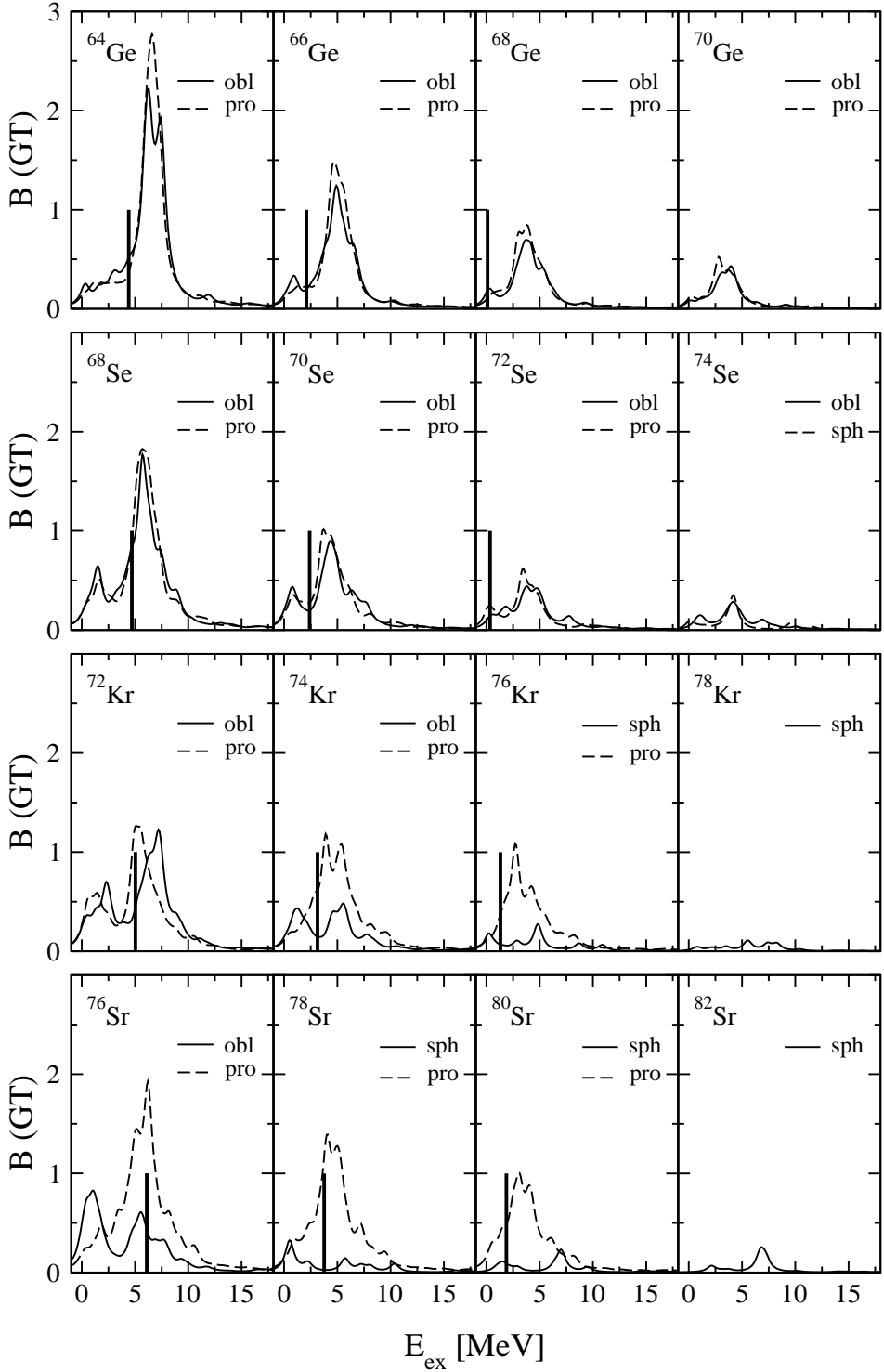}
\caption{Gamow-Teller strength distributions 
[$g_{A}^{2}/4\pi $] as a function of the excitation energy of the daughter 
nucleus [MeV]. The results are for the force SG2 in QRPA and
for the various shapes of the isotopes $^{64,66,68,70}$Ge,
$^{68,70,72,74}$Se, $^{72,74,76,78}$Kr, and $^{76,78,80,82}$Sr.
Vertical lines indicate experimental $Q_{EC}$ values
(see Table 1 for the theoretical $Q_{EC}$ values).}
\end{figure}

\vfill\eject

\begin{figure}[t]
\epsfig{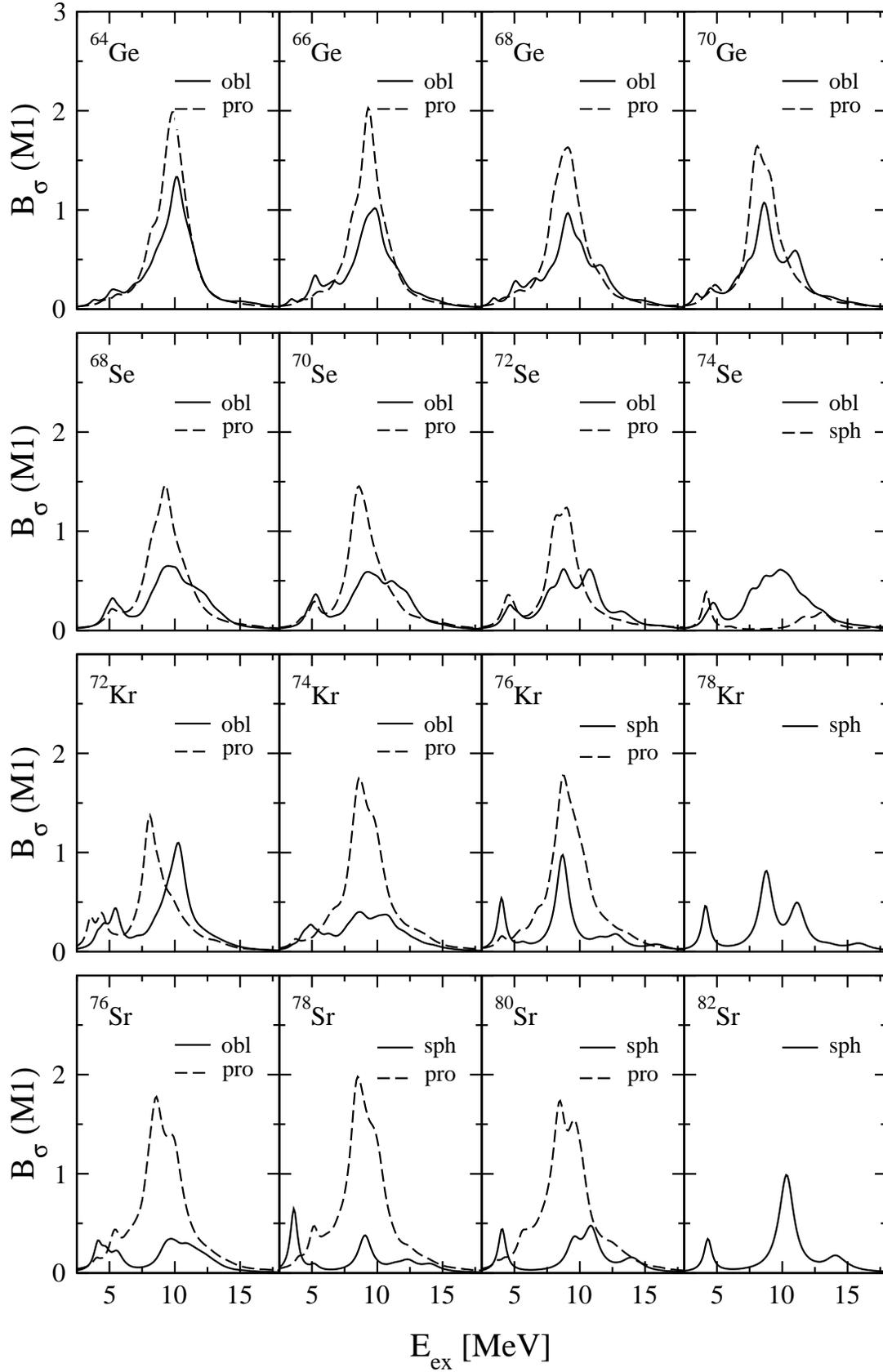}
\caption{Spin $M1$ strength distributions [$\mu_N^2$] calculated in
QRPA with the force SG2. The results are 
for the various shapes of the isotopes $^{64,66,68,70}$Ge,
$^{68,70,72,74}$Se, $^{72,74,76,78}$Kr, and $^{76,78,80,82}$Sr.}
\end{figure}

\vfill\eject

\begin{table}[t]

{\bf Table 1.} Results for $Q_{EC}$ values, half-lives $(T_{1/2})$, and 
GT strength summed up to  $Q_{EC}$ energies $(\sum_{Q_{EC}})$. The results 
correspond to QRPA calculations performed with the Skyrme force SG2 in four
isotopic chains and are calculated for the various equilibrium shapes in
each nucleus. Experimental values for $Q_{EC}$ and $T_{1/2}$ are from 
\cite{audi}.
\vskip 0.5cm
\begin{center}
\begin{tabular}{llccccccc}
&  & \multicolumn{2}{c}{$Q_{EC}$} &  & \multicolumn{2}{c}{$T_{1/2}$} &  & $%
\sum_{Q_{EC}}B(GT)$ \\ 
&  & exp & th &  & exp & th &  &  \\ \hline
&  & \multicolumn{1}{l}{} & \multicolumn{1}{l}{} &  & \multicolumn{1}{l}{} & 
\multicolumn{1}{r}{} &  &  \\ 
$^{64}$Ge & obl & \multicolumn{1}{l}{4.41} & \multicolumn{1}{r}{ 4.3 } &  & 
\multicolumn{1}{l}{63.7 s} & \multicolumn{1}{r}{84.5 s} &  & 0.7 \\ 
& pro & \multicolumn{1}{l}{} & \multicolumn{1}{r}{ 4.1} &  & 
\multicolumn{1}{l}{} & \multicolumn{1}{r}{167.0 s} &  & 0.5 \\ 
$^{66}$Ge & obl & \multicolumn{1}{l}{2.10} & \multicolumn{1}{r}{2.3} &  & 
\multicolumn{1}{l}{2.3 h} & \multicolumn{1}{r}{1.6 h} &  & 0.3 \\ 
& pro & \multicolumn{1}{l}{} & \multicolumn{1}{r}{2.2} &  & 
\multicolumn{1}{l}{} & \multicolumn{1}{r}{3.1 h} &  & 0.2 \\ 
$^{68}$Ge & obl & \multicolumn{1}{l}{0.11} & \multicolumn{1}{r}{0.2} &  & 
\multicolumn{1}{l}{271 d} & \multicolumn{1}{r}{198 d} &  & 0.1 \\ 
& pro & \multicolumn{1}{l}{} & \multicolumn{1}{r}{0.3} &  & 
\multicolumn{1}{l}{} & \multicolumn{1}{r}{100 d} &  & 0.0 \\ 
\hline 
$^{68}$Se & obl & \multicolumn{1}{l}{4.70} & \multicolumn{1}{r}{4.4} &  & 
\multicolumn{1}{l}{35.5 s} & \multicolumn{1}{r}{77.2 s} &  & 1.1 \\ 
& pro & \multicolumn{1}{l}{} & \multicolumn{1}{r}{4.5} &  & 
\multicolumn{1}{l}{} & \multicolumn{1}{r}{66.4 s} &  & 0.9 \\ 
$^{70}$Se & obl & \multicolumn{1}{l}{2.40} & \multicolumn{1}{r}{2.5} &  & 
\multicolumn{1}{l}{41.1 m} & \multicolumn{1}{r}{38.8 m} &  & 0.5 \\ 
& pro & \multicolumn{1}{l}{} & \multicolumn{1}{r}{2.7} &  & 
\multicolumn{1}{l}{} & \multicolumn{1}{r}{33.5 m} &  & 0.5 \\ 
$^{72}$Se & obl & \multicolumn{1}{l}{0.34} & \multicolumn{1}{r}{0.8} &  & 
\multicolumn{1}{l}{8.4 d} & \multicolumn{1}{r}{3.3 d} &  & 0.1 \\ 
& pro & \multicolumn{1}{l}{} & \multicolumn{1}{r}{1.3} &  & 
\multicolumn{1}{l}{} & \multicolumn{1}{r}{0.3 d} &  & 0.2 \\ 
\hline
$^{72}$Kr & obl & \multicolumn{1}{l}{5.04} & \multicolumn{1}{r}{5.0} &  & 
\multicolumn{1}{l}{17.2 s} & \multicolumn{1}{r}{21.4 s} &  & 1.2 \\ 
& pro & \multicolumn{1}{l}{} & \multicolumn{1}{r}{5.2} &  & 
\multicolumn{1}{l}{} & \multicolumn{1}{r}{13.6 s} &  & 1.9 \\ 
$^{74}$Kr & obl & \multicolumn{1}{l}{3.14} & \multicolumn{1}{r}{3.4} &  & 
\multicolumn{1}{l}{11.5 m} & \multicolumn{1}{r}{8.7 m} &  & 0.7 \\ 
& pro & \multicolumn{1}{l}{} & \multicolumn{1}{r}{3.5} &  & 
\multicolumn{1}{l}{} & \multicolumn{1}{r}{12.4 m} &  & 0.6 \\ 
$^{76}$Kr & sph & \multicolumn{1}{l}{1.31} & \multicolumn{1}{r}{1.7} &  & 
\multicolumn{1}{l}{14.8 h} & \multicolumn{1}{r}{4.1 h} &  & 0.2 \\ 
& pro & \multicolumn{1}{l}{} & \multicolumn{1}{r}{1.2} &  & 
\multicolumn{1}{l}{} & \multicolumn{1}{r}{38.0 h} &  & 0.1 \\ 
\hline
$^{76}$Sr & obl & \multicolumn{1}{l}{6.10} & \multicolumn{1}{r}{5.9} &  & 
\multicolumn{1}{l}{8.9 s} & \multicolumn{1}{r}{3.2 s} &  & 2.1 \\ 
& pro & \multicolumn{1}{l}{} & \multicolumn{1}{r}{5.8} &  & 
\multicolumn{1}{l}{} & \multicolumn{1}{r}{10.9 s} &  & 2.3 \\ 
$^{78}$Sr & sph & \multicolumn{1}{l}{3.76} & \multicolumn{1}{r}{4.3} &  & 
\multicolumn{1}{l}{2.7 m} & \multicolumn{1}{r}{1.3 m} &  & 0.4 \\ 
& pro & \multicolumn{1}{l}{} & \multicolumn{1}{r}{3.1} &  & 
\multicolumn{1}{l}{} & \multicolumn{1}{r}{19.9 m} &  & 0.6 \\ 
$^{80}$Sr & sph & \multicolumn{1}{l}{1.87} & \multicolumn{1}{r}{1.8} &  & 
\multicolumn{1}{l}{1.8 h} & \multicolumn{1}{r}{56.0 h} &  & 0.1 \\ 
& pro & \multicolumn{1}{l}{} & \multicolumn{1}{r}{1.6} &  & 
\multicolumn{1}{l}{} & \multicolumn{1}{r}{6.4 h} &  & 0.2\\
\end{tabular}
\end{center}
\end{table}

\vfill \eject

\begin{table}[t]
{\bf Table 2.} Comparison of the GT strengths contained below some given 
excitation energy between experimental measurements (\cite{exp72} for $^{72}$Kr, 
\cite{exp76} for $^{76}$Sr) and theoretical calculations.
\vskip 0.5cm
\begin{center}
\begin{tabular}{lccc}

& exp & oblate & prolate \\ 
\hline
$^{72}$Kr ($E_{ex} \leq $ 1.836 MeV) & 0.5 (1) & 0.5 & 0.8 \\
$^{76}$Sr ($E_{ex} \leq $ 2.882 MeV) & 0.6 -- 0.8 & 1.3 & 0.6 \\
\end{tabular}
\end{center}
\end{table}

\end{document}